\documentclass[graybox]{svmult}

\usepackage{mathptmx}       
\usepackage{helvet}         
\usepackage{courier}        
\usepackage{type1cm}        
\usepackage{makeidx}         
\usepackage{graphicx}        
\usepackage{multicol}        
\usepackage[bottom]{footmisc}

\makeindex             

\begin{document}

\title*{A nebula in your computer: simulating the physics and chemistry of an H~{\sc ii} region}
\titlerunning{A nebula in your computer}
\author{Thomas G. Bisbas}
\institute{University College London, Department of Physics and Astronomy, Gower Place, London WC1E 6BT, U.K., \email{tb@star.ucl.ac.uk}}
%
%
\maketitle

\abstract{In this contribution we discuss about numerical modeling of nebulae. In particular we emphasize on the dynamical evolution of an H~{\sc ii} region and the chemical structure of a Photodissociation region. We do this using the Smoothed Particles Hydrodynamics code \texttt{SEREN} and the recently developed astrochemistry code \texttt{3D-PDR}, respectively. We show an example application by simulating a cometary globule using these two codes.}

\section{Introduction}
\label{sec:1}

Nebulae (H~{\sc ii} regions) are large regions consisting of ionised gas, particularly of hydrogen. The source of ionisation is usually a single or multiple massive stars emitting ultraviolet radiation with photons carrying more energy than the ionisation potential ($h\nu>13.6{\rm eV}$). A nebula is mainly structured by three different parts; the ionised region (number density of $n\le200\,{\rm cm}^{-3}$ and temperature of $T\simeq10^4\,{\rm K}$), the photodissociation region\footnote{known also as ``Photon Dominated Region''}  (PDR; $200 \le n \le 10^5\,{\rm cm}^{-3}$, $10-20\le T\le 10^4\,{\rm K}$), and the dark molecular region ($n\ge 10^5\,{\rm cm}^{-3}$, $T\simeq10-20\,{\rm K}$). Of particular interest are PDRs; they are ubiquitously present in the interstellar medium (ISM) consisting of predominantly neutral gas and dust illuminated by FUV radiation ($6\le h\nu\le 13.6{\rm eV}$) and they occur in any region of the ISM that is dense and cold enough to remain neutral but has too low column density to prevent the penetration of FUV photons. Over the past few decades, effort has been made to study numerically the physics and chemistry of nebulae. Due to computational speed and memory capacity issues, detailed three-dimensional simulations are a reality since the past 5-7 years only. Even nowadays the numerical codes are divided in two main categories; codes that study the dynamical evolution and codes that study the chemical structure. An integrated code that treats simultaneously detailed dynamics, UV propagation and chemistry offering a realistic temperature and therefore pressure structure in H~{\sc ii} regions is still lacking (although significant effort in this direction has been made by \cite{Glov10, Clar12, Hawo12}).


\section{Dynamical \& Chemical modeling}
\label{sec:2}

Several techniques have been proposed by various workers on dynamical modeling of H~{\sc ii} regions in adaptive mesh refinement codes and in smoothed particle hydrodynamics codes. \cite{Bisb09} have proposed a \texttt{HEALPix}-based \cite{Gors05} algorithm to simulate the propagation of the UV radiation in the ISM by invoking the \emph{on-the-spot} approximation \cite{Oste74}. This algorithm has been incorporated in the SPH code \texttt{SEREN} and it creates hierarchies of rays emanated spherically symmetric from the excited source. Each ray splits up to four child-rays wherever the resolution of the radiation transfer matches the resolution of the hydrodynamics, locally. This algorithm adopts two equations of state; an isothermal for the ionised region ($T=10^4\,{\rm K}$ everywhere), and a barotropic for the neutral region (see Eqn. 6 \cite{Bisb11}). The temperature between these two regions, and therefore where the PDR is located, is smoothed using a linear interpolation. Although this technique speeds up the thermodynamical calculations, it significantly limits the ability to study the chemistry of PDRs which is quite important to understand in order to explore the insights of the chemical structure and star formation in ionised nebulae. We thus need further tools to implement.

Perhaps the most challenging part of modeling a nebula is its chemistry. That is because one has to take into account a realistic three-dimensional treatment of the UV radiation (i.e. as obtained by \texttt{MOCASSIN} \cite{Erco03, Erco05}), self-shielding of individual species against the UV radiation, cooling and heating processes and a complicated network of reactions. Of particular interest is the three-dimensional modeling of PDRs which has been achieved in \texttt{3D-PDR} for the first time. \texttt{3D-PDR} solves the chemistry and the thermal balance self-consistently within a given three-dimensional cloud of arbitrary density distribution. It uses the chemical model features of the fully benchmarked one-dimensional code \texttt{UCL\_PDR} \cite{Bell06} and a ray-tracing scheme based on the \texttt{HEALPix} package to calculate the total column densities and thus to evaluate the attenuation of the FUV radiation into the region, and the propagation of the FIR/submm line emission out of the region. An iterative cycle is used to calculate the cooling rates using a three-dimensional escape probability method, and heating rates. At each element within the cloud, it performs a depth- and time- dependent calculation of the abundances for a given chemical network to obtain the column densities associated with each individual species. The iteration cycle terminates when the PDR has obtained thermodynamical equilibrium, in which the thermal balance criterion is satisfied i.e. the heating and cooling rates are equal to within a user-defined tolerance parameter. \texttt{3D-PDR} determines the relative abundances of a limited number of atomic and molecular species at each cloud element, by solving the time-dependent chemistry of a self-contained network of formation and destruction reactions. We use the UMIST database containing 33 species (including ${\rm e}^-$) and 320 reactions. We solve for steady-state chemistry (chemical evolution time set to $t=100\,{\rm Myr}$), although the code is able to follow the full time dependent evolution within the cloud.

This treatment of PDR chemistry offers a more accurate temperature profile in comparison with the approximations made in the dynamical simulations. However, due to the very high computational demanding, it is impossible to include such complicated UV and PDR calculations in a dynamical code, unless new techniques are implemented.

\section{Example: simulating a cometary globule}
\label{sec:4}

\begin{figure}[t]
\includegraphics[width=0.33\textwidth]{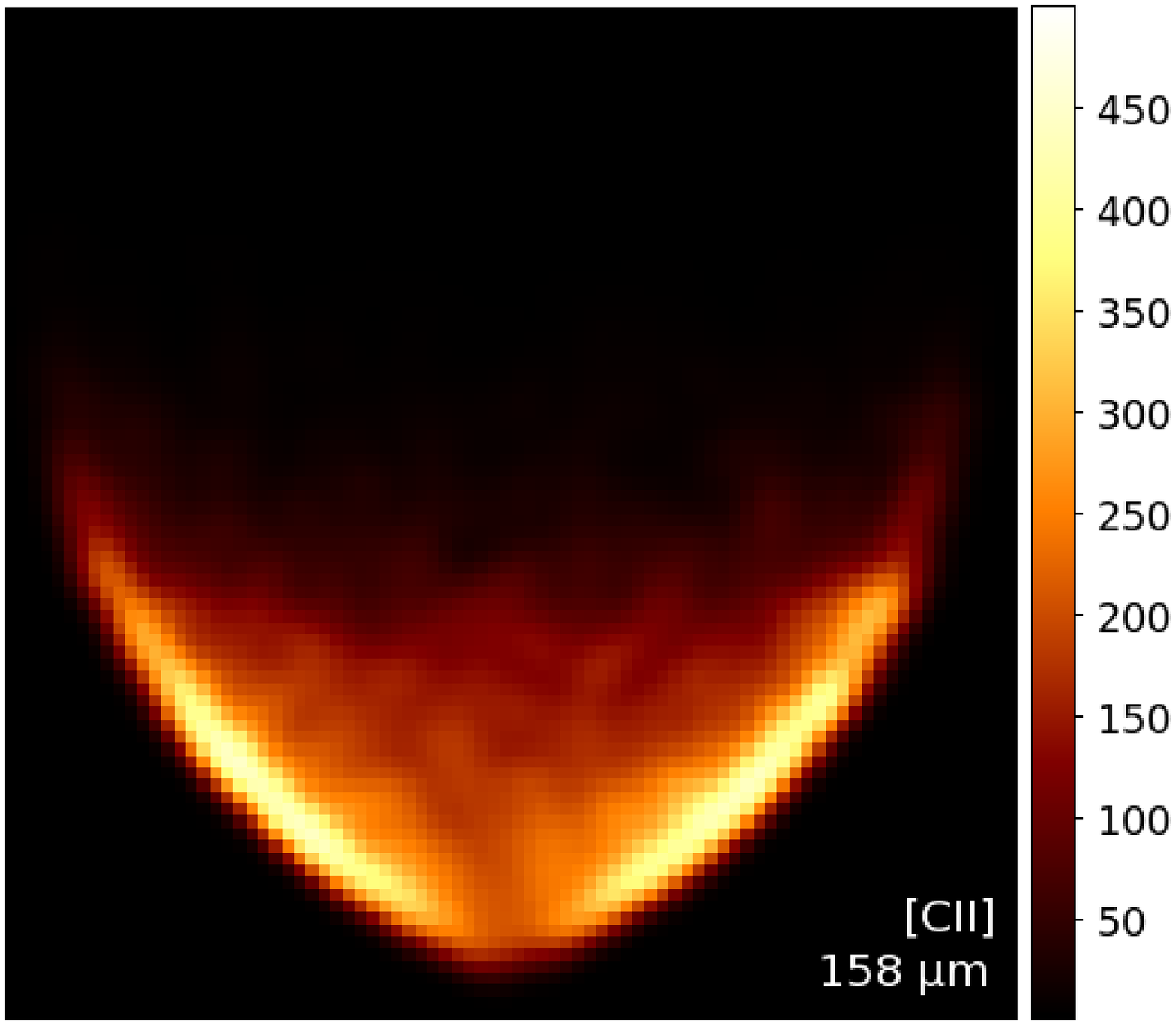}
\includegraphics[width=0.33\textwidth]{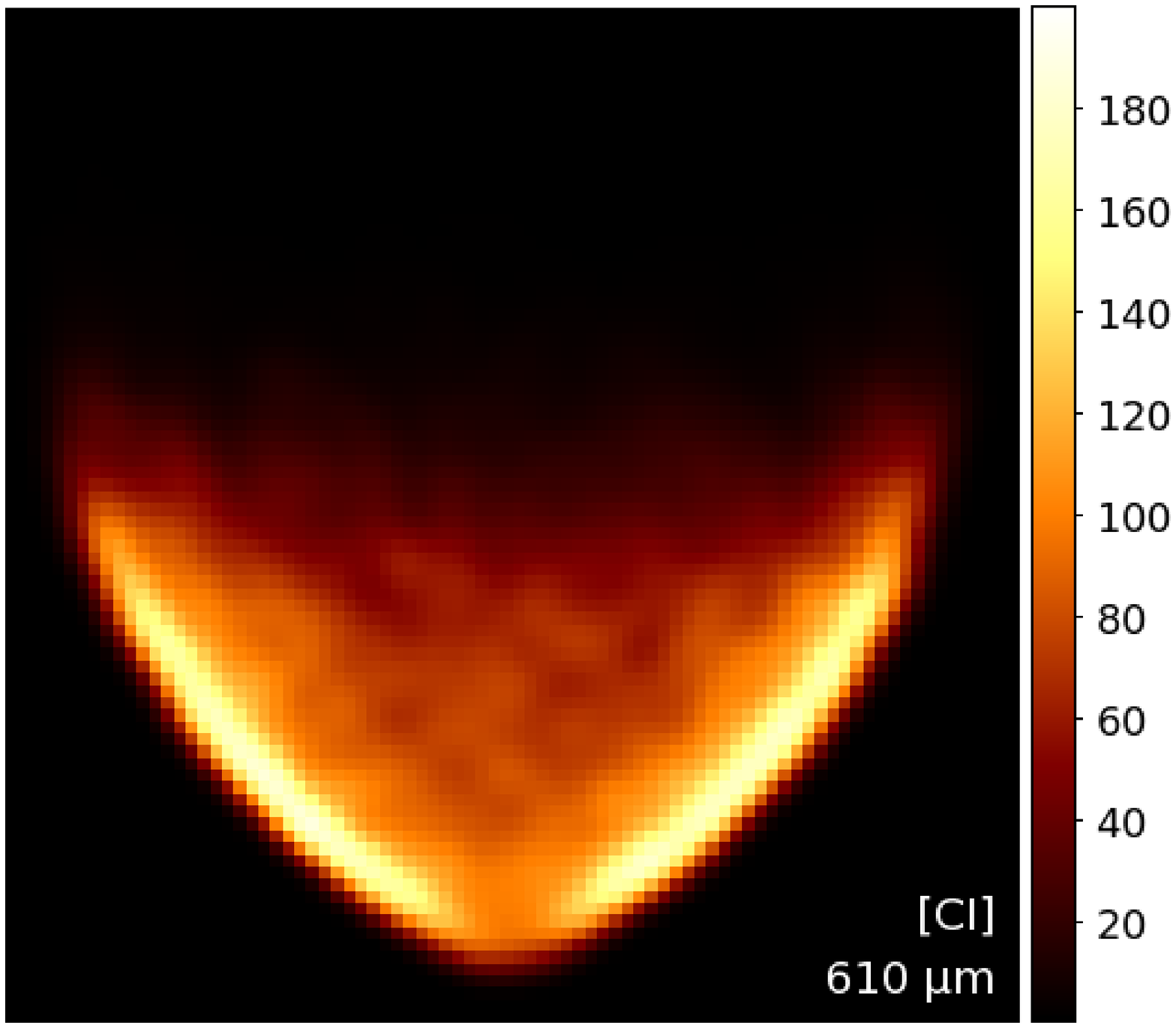}
\includegraphics[width=0.33\textwidth]{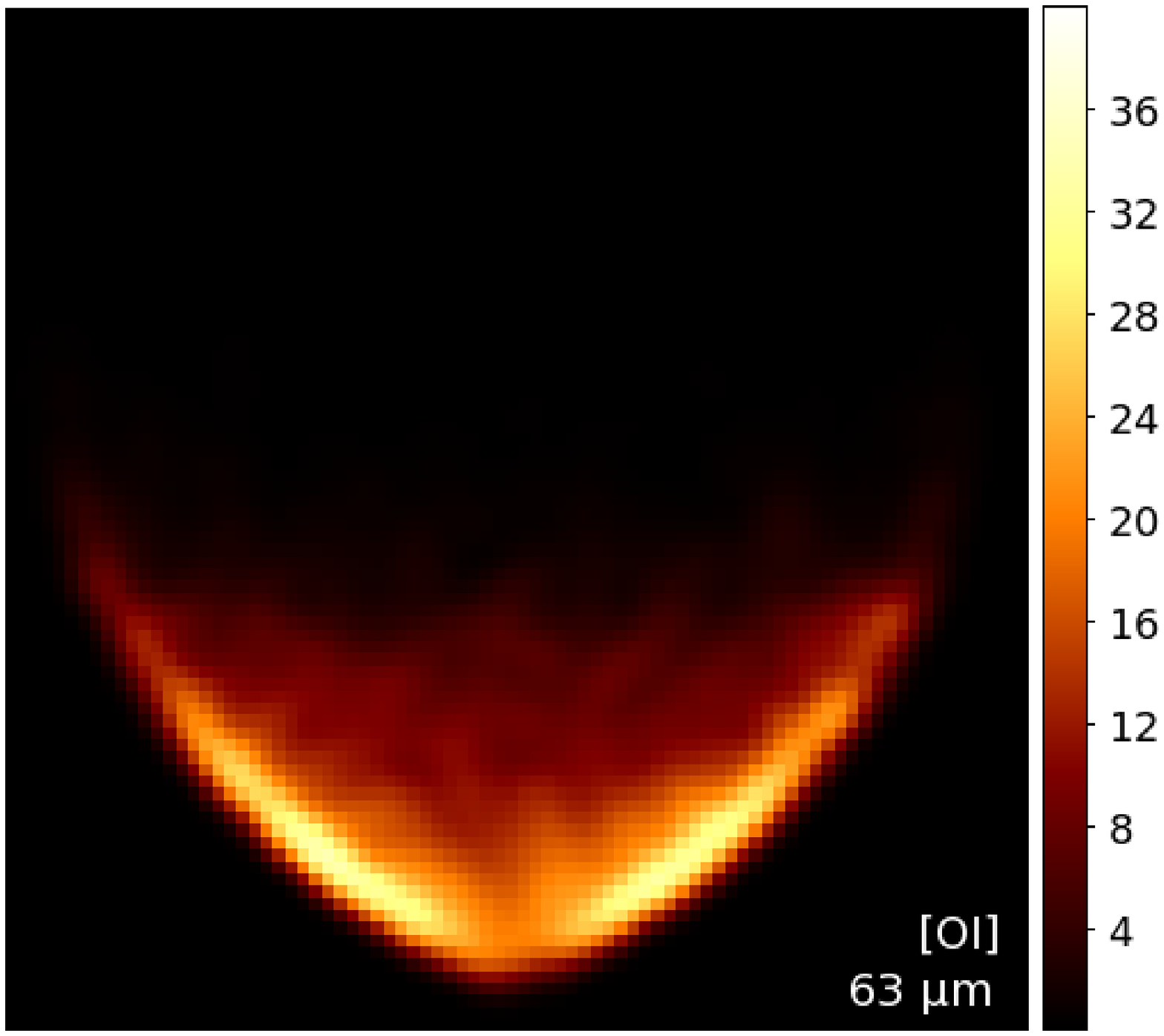}
\includegraphics[width=0.33\textwidth]{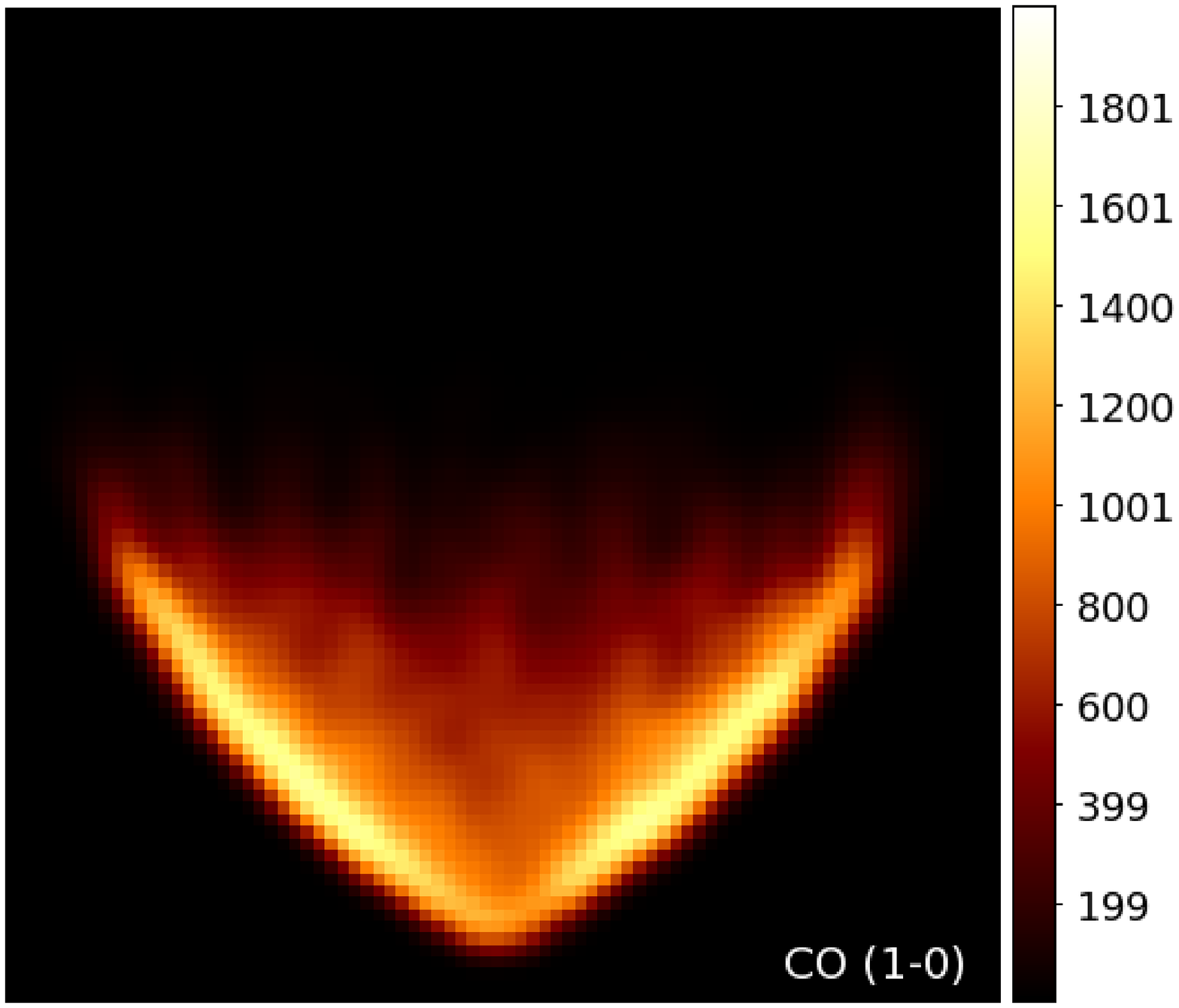}
\includegraphics[width=0.33\textwidth]{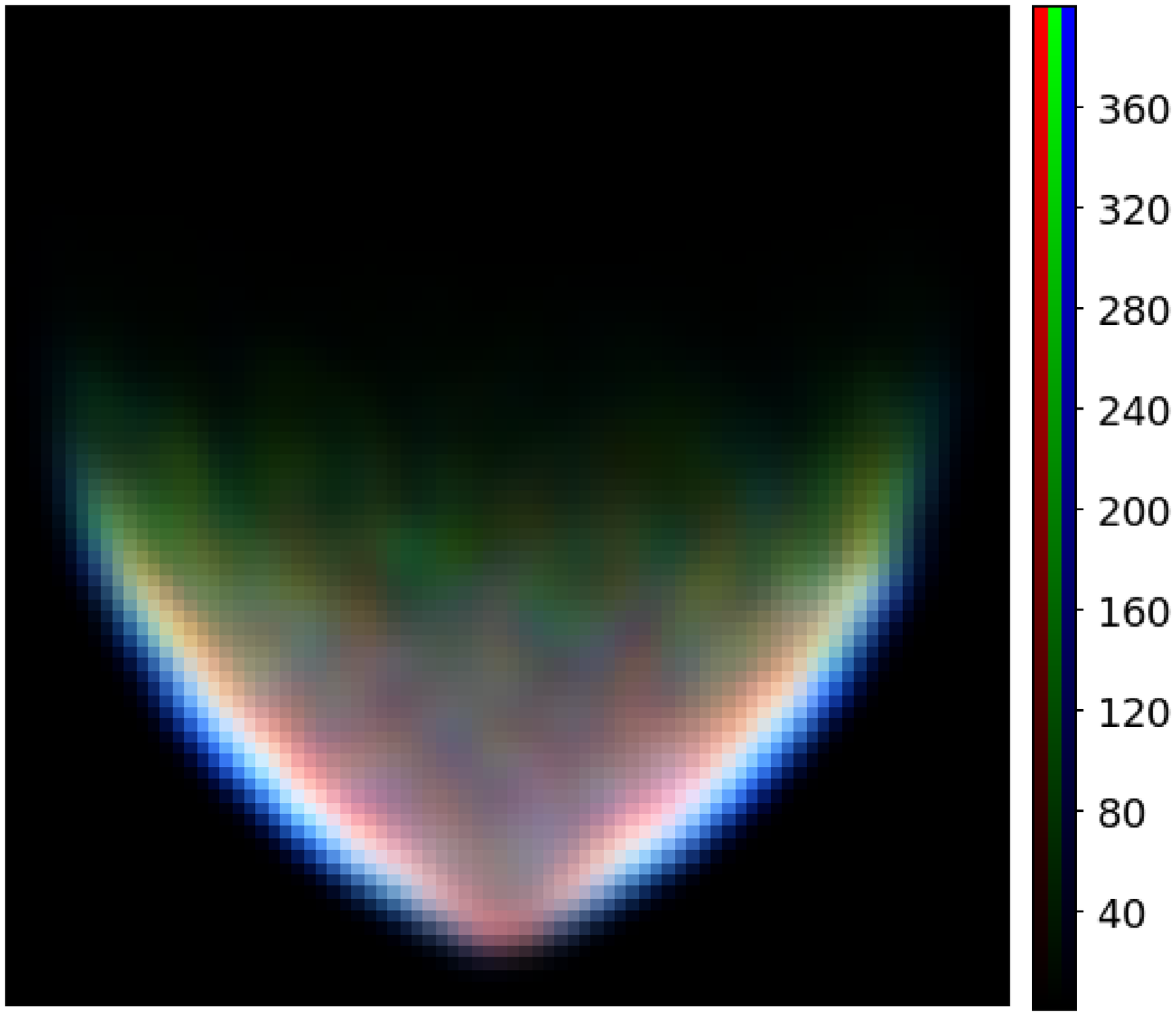}
\includegraphics[width=0.33\textwidth]{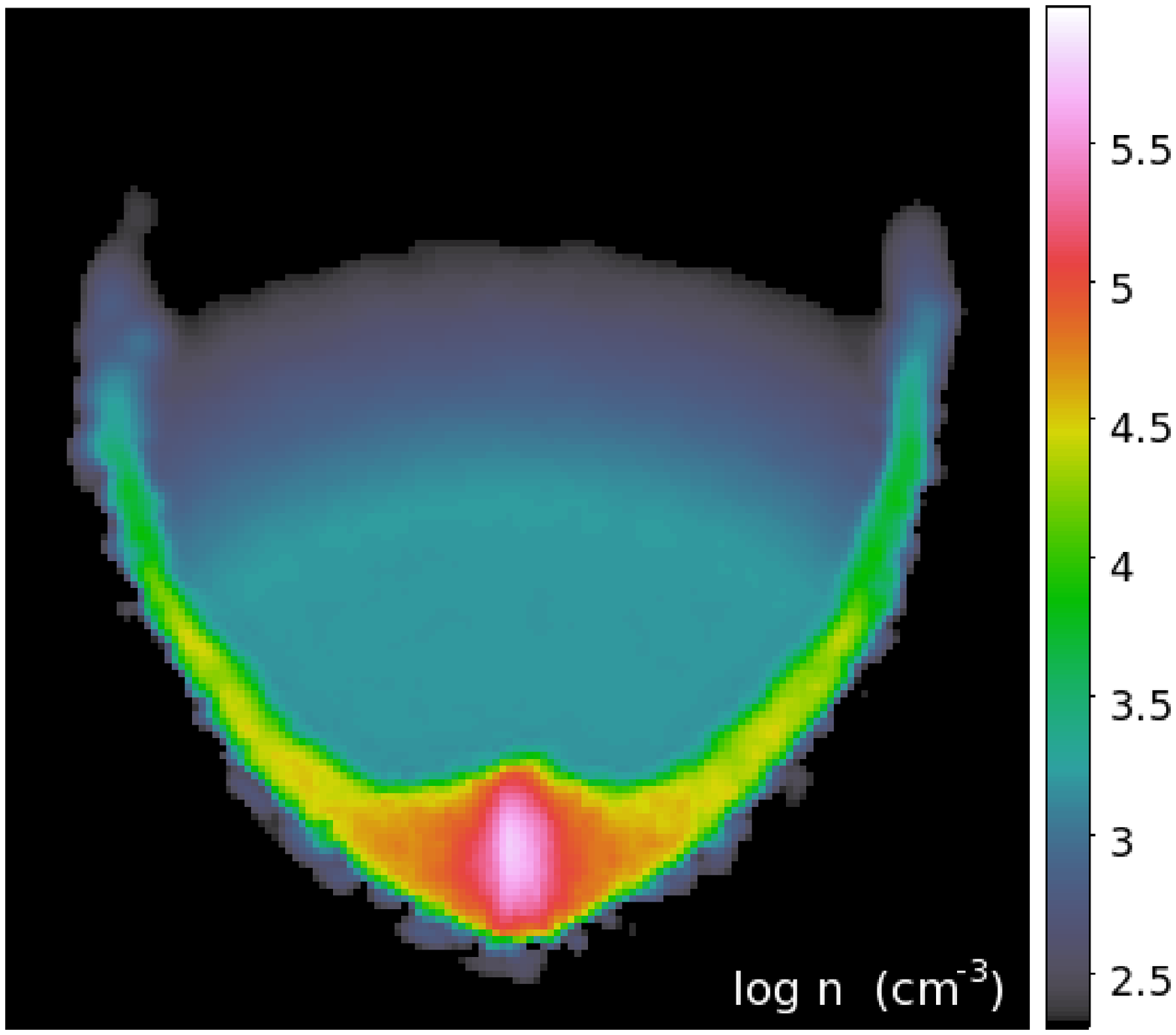}

\caption{ Top row from left to right: emission maps of [C~{\sc ii}] $158\,\mu{\rm m}$, [C~{\sc i}] $610\,\mu{\rm m}$ and [O~{\sc i}] $63\,\mu{\rm m}$. Bottom row from left to right: emission map of CO (1-0), RGB composite image with colour bar ratios of 8:1:2 for CO(1-0):[C~{\sc i}]:[C~{\sc ii}], cross-section density profile at $z=0\,{\rm pc}$.}
\label{fig:1}
\end{figure}

In this example we perform two different simulations using the codes described above. In the first simulation (SPH run) we use the \texttt{SEREN} code to evolve dynamically an initially uniform density spherical clump (radius $R=0.5\,{\rm pc}$, mass $M=20\,{\rm M}_{\odot}$, temperature $T=100\,{\rm K}$, consisting purely of atomic hydrogen) as it interacts with an external and approximately plane-parallel radiation field. In the second simulation (PDR run) we use the \texttt{3D-PDR} code in one of the snapshots selected from the SPH run in which we examine the PDR chemical structure.

In the SPH run, as the clump interacts with the external radiation field emitted spherically by an excited source (at a rate of $\dot{\cal N}_{_{\rm LyC}}=3.2\times10^{48}$ photons ${\rm s}^{-1}$, placed $D=3.5\,{\rm pc}$ away from the centre of the clump) the ionising radiation boils off its outer layers from the side in which the flux is impinging. The shock front that is formed compresses the remaining neutral gas turning it into a rod shape. However, the internal thermal pressure of the neutral gas is constantly increasing and this results in its re-expansion (see \S4.4 of \cite{Bisb09} for full discussion), which may lead to star formation \cite{Grit09, Bisb11}.

In the PDR run, we take a snapshot from the SPH run at $t=0.12\,{\rm Myr}$ and we use it as initial conditions in \texttt{3D-PDR}. A cross section plot of the density structure at that time is shown at the bottom right of Fig. \ref{fig:1} where the ionising radiation is impinging from bottom to top. The UV photon flux corresponds to a field strength of approximately $\chi=50\,{\rm Draines}$. The top three and the bottom left frames show the emission maps for the most dominant coolants. The middle frame at the bottom shows an RGB composite image of three different emission maps. From these maps we see that the species in the PDR are distributed smoothly and follow the density profile. The weakest emission is produced by [O~{\sc i}] $63\,\mu{\rm m}$ and the strongest by CO (1-0) implying that the molecular gas dominates over the atomic contribution. However, considering the transition frequencies for these maps, we find that the [O~{\sc i}] $63\,\mu{\rm m}$ and the [C~{\sc ii}] $158\,\mu{\rm m}$ lines are the dominant coolants. Full details and results are discussed in \cite{Bisb12}.

\section{Conclusions}
\label{sec:5}

We discussed about the dynamical and chemical modeling of H~{\sc ii} regions using the \texttt{SEREN} SPH code and the recently developed \texttt{3D-PDR} code respectively. We perform an example of a cometary globule and we show results from these two codes. Since we lack of algorithms treating detailed dynamical and chemical calculations in ionized regions, effort has to be made towards this direction.

\begin{acknowledgement}
TGB acknowledges support by STFC grant ST/H001794/1
\end{acknowledgement}

\end{document}